\documentclass{article}
\usepackage[T1]{fontenc}
\usepackage[utf8]{inputenc}
\usepackage{ismir} 
\usepackage{amsmath,cite,url}
\usepackage{graphicx}
\usepackage{color}
\usepackage{amssymb}
\title{CrowdioSet and PaRIRset: Two Datasets Towards Live Music Source Separation}

 \multauthor
   {Enric Gusó$^{1,2}$ \hspace{1cm} Xavier Serra$^1$ \hspace{1cm}}
   {
   $^1$ Music Technology Group, Universitat Pompeu Fabra, Barcelona, Spain\\
   $^2$ Eurecat, Centre Tecnològic de Catalunya, Barcelona, Spain\\
   {\tt\small enric.guso@upf.edu, xavier.serra@upf.edu}
   }

\def\authorname{E. Guso and X. Serra}

\usepackage[bookmarks=false,pdfauthor={\authorname},pdfsubject={\pdfsubject},hidelinks]{hyperref}

\begin{document}
\thispagestyle{empty}

\maketitle

\begin{abstract}

Most Music Source Separation (MSS) models do not generalize well to live music recordings because they are trained on studio recordings alone, disregarding the venue acoustics, the speaker system's response and audience noise. We propose to bridge this gap by providing and training a model on two novel datasets. First, we present CrowdioSet: a noise dataset comprising 4800 real ambience tracks from Freesound and synthetic sing-alongs for the vocals in MUSDB18 and MOISESDB datasets, generated from zero-shot singing voice conversions. CrowdioSet enables effective audio denoising for live recordings, resulting in superior separation both in objective and subjective evaluations. Second, we introduce PaRIRset, a stereo impulse response dataset captured across 40  professional concert venues using a microphone array. Our results show that  adding PaRIRset RIRs increases the performance of a MSS model compared to using real RIRs from speech enhancement tasks alone. We make the examples, code, model weights, PaRIRset, and CrowdioSet freely available to the public.

\end{abstract}

\section{Introduction}\label{sec:introduction}

Music Source Separation (MSS) has advanced rapidly since the standardization of the task \cite{rafii2017musdb18, stoter2019open}. Architectures such as \cite{rouard2023htdemucs, luo2023bsrnn, tong2024scnet} and community challenges \cite{fabbro2024sdx23, zang2026msr, mitsufuji2022music} have progressively improved separation quality or expanded the scope of the task;  from generative approaches \cite{plaja2025generating}, through revisiting evaluation metrics \cite{sheridan2025perceptual, jaffe2025musical}, to lightweight models \cite{hung2025moises, venkatesh2024real}, and moving beyond the fixed \textit{vocals, drums, bass, other} taxonomy towards universal \cite{postolache2023adversarial} and prompt-conditioned separation \cite{shi2025samaudio}.

However, most MSS models are trained exclusively on studio recordings where sources mix linearly---an assumption that breaks down for live music, where venue acoustics, loudspeaker response, and audience noise reshape the signal. The only prior works in live settings are \cite{plaja2023carnatic}, which focuses on Carnatic music, and \cite{kandpal2022musicenhancement}, which addresses music enhancement without performing separation.

\begin{figure}[h!]
  \centering
  \includegraphics[trim={1.2cm 0.6cm 0.5cm 0.3cm}, clip,width=0.85\linewidth]{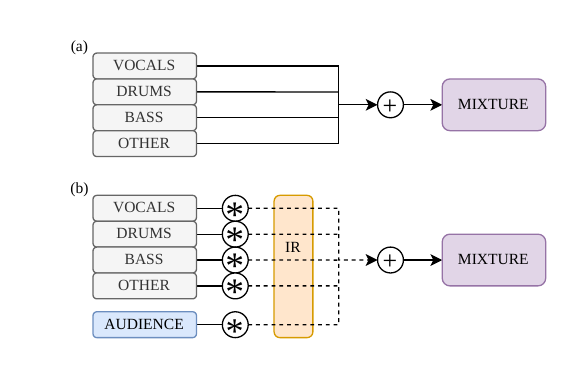}
  \caption{We propose to extend (a) Traditional Music Source Separation to (b) Live Music Source Separation, by adding \textit{audience} and reverberation before mixing.}
  \label{fig:paradigm}
\end{figure}

Beyond the music domain, in speech separation WHAM! \cite{wichern2019wham} and WHAMR! \cite{maciejewski2020whamr} datasets successfully bridged an analogous gap by augmenting the clean WSJ0-2mix corpus with ambient noise and reverberation respectively. In this work we follow the same rationale for MSS, extending traditional MSS datasets to the Live Music Source Separation as depicted in Figure~\ref{fig:paradigm}.

Our approach specifically targets \emph{audience-side} recordings where audio quality is degraded by venue acoustics and crowd noise, but other use cases could benefit from it; the Cadenza Challenges \cite{roa2025first} have recently demonstrated how MSS can help hearing-aid users, which could be extended to the live setting, as well as other MIR tasks such as fingerprinting. Overall, we have made the following contributions:

\begin{itemize}
    \item \textbf{CrowdioSet}: synthetic \textit{audience} stems, comprising 4800 real noise tracks from Freesound\cite{anastasopoulou2025freesound} and a preliminary pipeline for generating \textit{sing-alongs} for every \textit{vocals} stem in MUSDB18HQ \cite{rafii2017musdb18} and MOISESDB \cite{pereira2023moisesdb} datasets, generated from a zero-shot singing voice conversion model HQ-SVC \cite{bai2026hq}.
    \item \textbf{PaRIRset}: 40 measured multichannel impulse responses from professional concert venues, augmented to 2200 stereo impulses.
    \item \textbf{SCNet model} \cite{tong2024scnet}: we have re-trained an open-source MSS model on the augmented datasets.
\end{itemize}

\section{CrowdioSet Audience Dataset}\label{sec:crowdioset}
We have identified three main categories of sounds through informal listening to concert recordings from the audience: (i) \textit{ambiences}, namely the room tone and background chatter; (ii) \textit{events}, such as applause, cheering, and shouts of approval or disapproval; and (iii) \textit{sing-alongs}, sections in which the audience sings the lyrics along with the performer. CrowdioSet is split into two parts: a selection of Freesound recordings for \textit{ambiences} and \textit{events}, which are the focus of our contribution, and a preliminary generative pipeline for \textit{sing-alongs}, which we release as a first step towards covering this third category (see Section~\ref{sec:limitations_future}).

\subsection{Ambiences and Events}\label{subsec:ambiences_events}

We have downloaded 11375 audio files from Freesound using the queries \textit{crowd}, \textit{audience}, \textit{cheering}, \textit{applause}, \textit{chatter}, and \textit{protest}, restricting results to files shorter than eight minutes and to licenses CC0, CC-BY 3.0/4.0, CC-BY-NC 3.0/4.0, and Sampling+ 1.0. Per-file attribution and license metadata are released alongside the dataset. Each file has been resampled to 44.1\,kHz, 16-bit stereo WAV. We have manually discarded recordings that did not plausibly occur at a concert (e.g., traffic, nature, or machine sounds) and classified the remaining 4819 files as either \textit{ambiences} or \textit{events} as shown in Table~\ref{tab:crowdioset_queries}.

\begin{table}[h]
  \centering
  \small
  \begin{tabular}{l|cc|cc}
    & \multicolumn{2}{c|}{\textit{ambiences}} & \multicolumn{2}{c}{\textit{events}} \\
     & \#files & hours & \#files & hours \\
    \hline
    crowd     & 1997 & 73.26 & 1398 & 10.40 \\
    audience  &  262 &  8.91 &  690 &  3.65 \\
    chatter   &  105 &  4.01 &   10 &  0.04 \\
    cheering  &   15 &  0.72 &  119 &  1.10 \\
    applause  &    0 &  0.00 &  182 &  1.70 \\
    protest   &   30 &  1.21 &   11 &  0.13 \\
    \hline
    Total     & 2409 & 88.11 & 2410 & 17.02 \\
  \end{tabular}
  \caption{CrowdioSet query distribution across the \textit{ambiences} and \textit{events} categories (105.12 hours total).}
  \label{tab:crowdioset_queries}
\end{table}

The majority of the retained files (91.98\%: 58.79\% CC0, 33.19\% CC-BY across both 3.0 and 4.0) are usable in commercial applications, while 8.02\% are CC-BY-NC and would require retraining without them for commercial use. To match MUSDB18HQ, we reserve 50 files per category for testing and 14 per category for validation.

\subsection{Synthetic Sing-Alongs}\label{subsec:singalongs}

\begin{figure}[h]
  \centering
  \includegraphics[width=\linewidth]{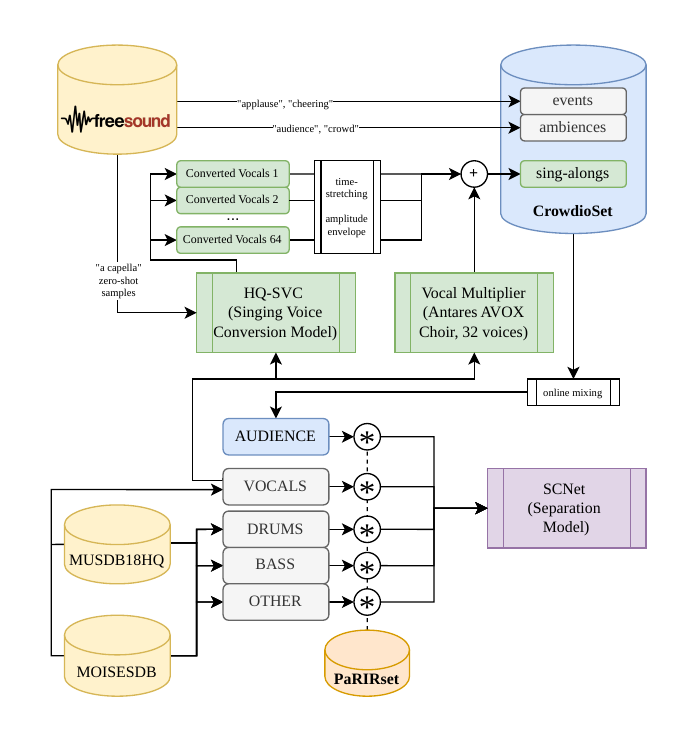}
  \caption{Data generation pipeline. Sing-along generation of every \textit{vocals} stem by using the HQ-SVC model and the AVOX Choir audio effect (green), and online \textit{audience} assembly with \textit{ambiences} and \textit{events} (blue).}
  \label{fig:singalong_pipeline}
\end{figure}

As shown in Figure~\ref{fig:singalong_pipeline}, we have synthesized preliminary sing-alongs for every one of the $N=384$ \textit{vocals} stems in MUSDB18HQ and MOISESDB by combining two techniques. First, we have applied the Antares AVOX Choir plugin (an effect that combines vibrato, detuning and delay) in its 32-voice configuration to each \textit{vocals} stem. Second, we have sourced a pool of $I=200$ vocal samples from Freesound using the queries \textit{a cappella} and \textit{vocals}, and used each to convert every \textit{vocals} stem using the pre-trained zero-shot voice conversion model HQ-SVC~\cite{bai2026hq}. We have generated $I/2$ conversions preserving the original $f_0$ contour from the stem ---yielding conversions perfectly in tune but with similar timbre--- and $I/2$ conversions adjusting it to the range of the zero-shot sample ---producing different timbre but less in tune.

Since realistic audience sing-alongs should feature voices that are timbrally diverse yet remain loosely coupled to the melody of the lead vocal, we have ranked the conversions by a combined similarity score that balances these two criteria. For each candidate $v_i$ and the original vocals stem $v_o$, we have computed a chroma mean absolute error $d_i = \text{MAE}\!\left(\Phi(v_i),\, \Phi(v_o)\right)$, where $\Phi(\cdot)$ denotes the chromagram using 12 bins, 2048-sample Hann window and 512 hop size, and a singer-identity similarity $s_i = \cos\!\left(\,\Psi(v_i),\, \Psi(v_o)\right)$, where $\Psi(\cdot)$ is the singer embedding from \cite{torres2023singer}. Small $d_i$ implies agreement on pitch content (in tune with the lead vocal), while small $s_i$ implies distinct vocal identity. We have standardized both across $N \cdot I$ pools as $\tilde{d}_i = (d_i - \mu_d)/\sigma_d$ and $\tilde{s}_i = (s_i - \mu_s)/\sigma_s$, and have sorted $i^\star$ by $\arg\min (\tilde{d}_i + \tilde{s}_i)$.

Then we have taken the best 64 candidates and have processed each one independently: we have approximated a time map of the syllables by applying onset detection \cite{mcfee2025librosa} on $v_o$. We have changed the amplitude of each syllable by drawing gains from $\mathcal{U}(0.3, 1)$ and each syllable's duration by adding delays  $\sim \mathcal{U}(0, 0.5)\,\text{s}$ to each entry of a time-stretching time map. Every individual voice has been summed into a stereo mix after applying another delay $\sim \mathcal{U}(0, 0.3)\,\text{s}$ to the whole voice. We have randomly panned each voice to the left and right stereo channels by independent gains $\sim \mathcal{U}(0, 1)$. Finally, the HQ-SVC mix has been mixed by ear with the one from the AVOX Choir forming the \textit{sing-along}.

We provide separate \textit{sing-alongs} for every MOISESDB song and also for the training set of the MUSDB18HQ. During training, we have generated the \textit{audience} stems on the fly in the data loader. Starting from silence, we have independently added each component with a random gain and probability: for \textit{sing-alongs}, \textit{ambiences} and \textit{events}, we have included $v$, $a$, and $e$ with probabilities $p_v = p_a = p_e = 0.5$ respectively; and applied gains $g_v$, $g_a$ and $g_e$ from $\sim \mathcal{U}(0,1)$. This stochastic mixing exposes the model to a wide variety of audience conditions without requiring pre-generated training mixes. For the MUSDB18HQ test and validation sets, we have provided fixed, manually pre-generated \textit{audience} mixes instead, to ensure reproducible evaluation.

\section{PaRIRset Reverberation Dataset}\label{sec:parirset}

On top of the audience noise, we propose to simulate the acoustics of modern concerts. As in \cite{kandpal2022musicenhancement}, we have started from RIR collections commonly used in Speech Enhancement (SE): the ACE Challenge dataset \cite{eaton2015ace}, the MIT IR Survey \cite{traer2016statistics}, and the SLR28 corpus \cite{ko2017study}. However, these RIRs are limited for our purposes in two ways: first, the rooms captured (offices, classrooms, lecture halls) bear little resemblance to concert venues; second, modern live sound is delivered through a Public Address (PA) system with separate left and right loudspeaker arrays for stereo reproduction, a configuration absent from SE datasets. We therefore measured our own set of impulse responses, which we call \emph{Public Address Room Impulse Response Set} (PaRIRset)--- the first dataset of real RIRs captured in professional concert venues with PA systems.

\begin{figure}[h]
  \centering
  \includegraphics[trim={1cm 2cm 0.5cm 3cm},clip,width=0.9\linewidth]{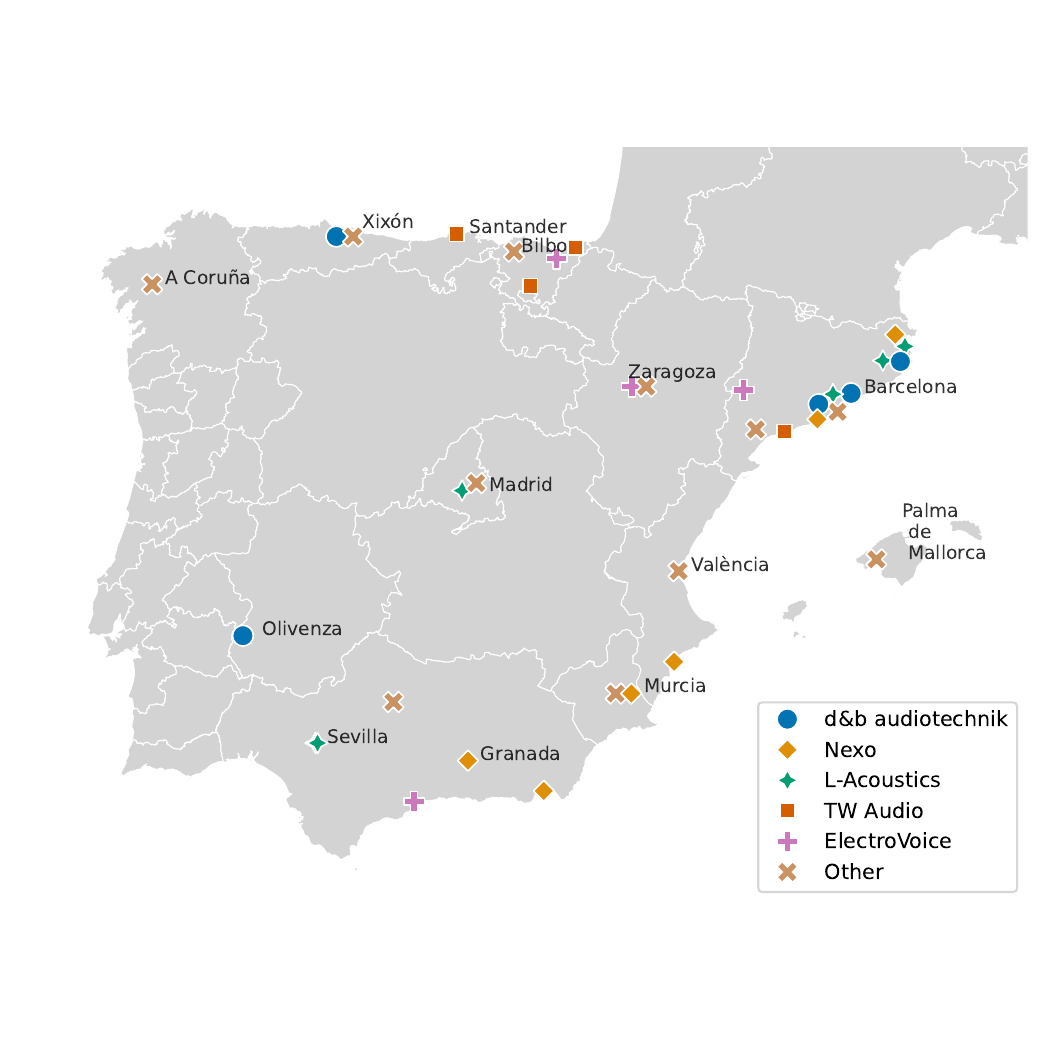}
  \caption{PaRIRset venue locations and the manufacturer of each venue's PA.}
  \label{fig:map}
\end{figure}

PaRIRset covers 40 professional concert venues spanning a wide geographical area, as shown in Figure~\ref{fig:map}. More importantly, the dataset also reflects diversity in PA manufacturers (d\&b~audiotechnik, L-Acoustics, Nexo, TW~Audio, Electro-Voice, among others), also covering different PA sizes (from small concerts for 300 people to arena-sized PAs). We provide more details of each venue and PA in the data repository, including calibration values, venue websites, whether the RIR was measured indoors or outdoors, and the specific PA model and element count (how many elements of each speaker model formed each side or each subsystem of the PA). Technically, 9 of the 40 PaRIRset measurements correspond to outdoor PA systems. We make the raw mono and multichannel recordings available at $f_s=48$\,kHz sampling rate, as well as the processed, augmented stereo RIRs used for training, downsampled to $44.1$\,kHz to match the rest of training datasets.

\subsection{RIR Measurement Methodology}\label{subsec:measurement}

\begin{figure}[h]
  \centering
  \includegraphics[trim={2cm 5.1cm 2cm 2cm},clip,width=0.65\linewidth]{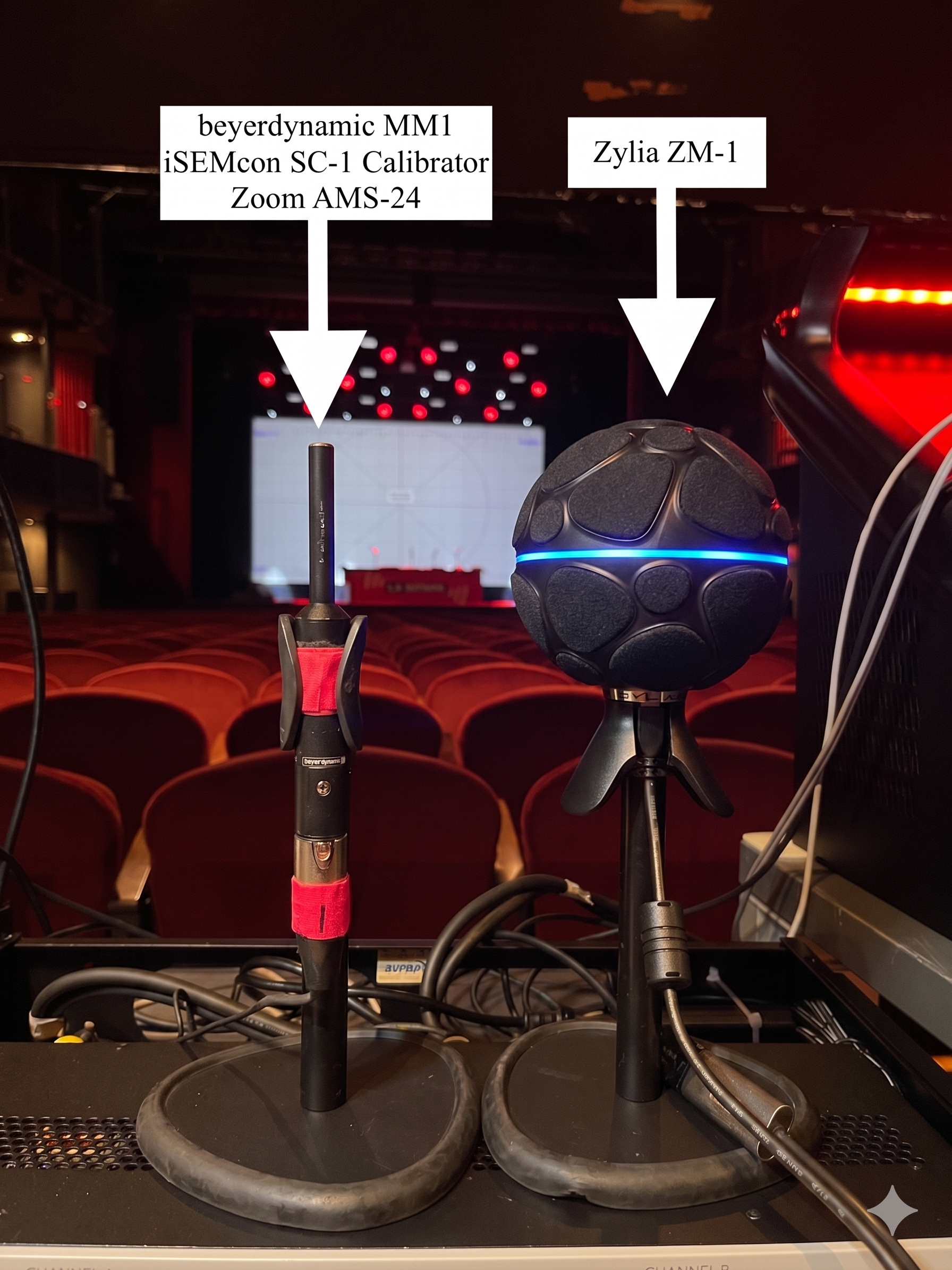}
  \caption{PaRIRset recording setup at the Front of House (FOH) position, near the sound engineer's mixer.}
  \label{fig:setup}
\end{figure}

Access to professional venues and their PA systems is limited, so we have had to limit our measurements to performing a single exponential sine sweep per PA side (left and right) at each venue. The measurement point has been placed next to the mixing console, commonly known as the Front of House (FOH) position, which should be the most representative listening position in the room. As shown in Figure~\ref{fig:setup}, we used a Zylia ZM-1 microphone array alongside a beyerdynamic MM1 measurement microphone connected via a Zoom AMS-24 interface.

RIRs have been captured using the exponential sine sweep method in \cite{farina2000simultaneous} with parameters $f_1 = 20$\,Hz, $f_2 = 20$\,kHz and a sweep duration of $T = 6$\,s. After deconvolution, two post-processing steps have been applied. First, noise has been reduced by applying a rectangular window to each RIR when its amplitude envelope fell below the noise floor: we have retained 100\,ms before the direct sound, applied a 50\,ms linear fade-in at the start, and a 50\,ms fade-out before the noise floor onset. Second, since the FOH position is not always equidistant from both PA sides, the left and right RIRs have been time-aligned via the cross-correlation method. Using the beyerdynamic MM1 as a flat-response reference, we additionally applied a second-order high-shelf filter to the Zylia-derived RIRs ($f_0 = 3.5$\,kHz, gain $= -4$\,dB, $Q = 0.5$).

\subsection{RIR Data Augmentation Strategies}\label{subsec:augmentation}

We have explored four strategies to determine how best to augment the 40 RIR measurements. First we have used the beyerdynamic RIRs and performed the following basic augmentation $(D_1)$: given the concatenation operator $\circ$ and gains $g = \{-6, -1.5, 0, 1.5, 3.5\}$\,dB, we have split each original RIR $h$ at the sample $argmax( | h | )+2 \cdot 10^{-3} \cdot f_s$ into direct sound $d_h$ and tail $t_h$ and modified polarity and gain of each part, i.e. $\{h, -h, d_h \circ (t_h \cdot g), -d_h \circ (t_h \cdot g), d_h \circ (t_h\cdot g)\}$, augmenting each measurement 20 times. 

Then we have explored $(D_2)$: taking $D_1$ and adding the Zylia RIRs using the raw multichannel recordings from each of the 19 capsules ---each with its own small displacement, which we call $\text{zylia}_{caps}$; $(D_3)$ generating 19 virtual microphones with the array's proprietary beamforming algorithm (Zylia Studio) with random orientations and polar patterns, which we call $\text{zylia}_{beams}$; and $(D_4)$ permuting $d_h$ from one venue with $t_h$ of another, which we call $\text{PA}_{permute}$.

\begin{figure}[h]
  \centering
  \includegraphics[trim={2.4cm .8cm .2cm .2cm}, clip,width=\linewidth]{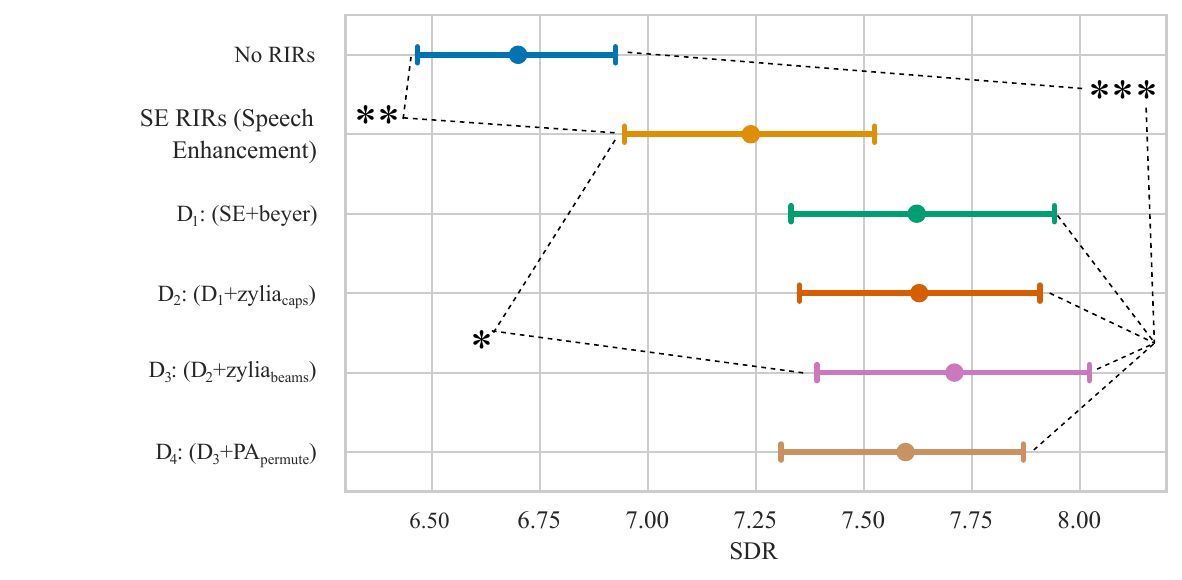}
  \caption{Ablation results for PaRIRset augmentation strategies, adding one set of augmented RIRs at a time. Significance markers from paired $t$-tests: * $p \leq 0.05$, ** $p \leq 0.01$, *** $p < 0.001$, and non-significant in absence. Error bars denote 95\% confidence intervals.}
  \label{fig:parirset_sdr}
\end{figure}

For assessing the different augmentation strategies, we have retrained a MSS SCNet~\cite{tong2024scnet} model from scratch on MUSDB18HQ, using all the default hyperparameters, but convolving all sources and targets with the RIRs, adding each set of augmented RIRs to the training data, one at a time. We have evaluated with 8 reserved venues, augmented with $D_1$ and forming PaRIRset's test set. In Figure~\ref{fig:parirset_sdr} we report the standard Signal to Distortion Ratio (SDR) across all sources on the test set. We also provide two augmentation baselines: using no reverberation at all, and convolving only with the RIRs from the SE datasets.

Results in Figure~\ref{fig:parirset_sdr} show that all configurations that include PaRIRset RIRs significantly outperform the no-reverb baseline. While $D_1$ --- $D_4$ do not differ substantially from each other, only $D_3$ achieves a statistically significant improvement over the SE-only baseline (paired $t$-test, $p = 0.027$). The $\text{PA}_{permute}$ strategy $D_4$ did not yield further gains, so we have discarded it. Henceforth, with PaRIRset we refer to the combination of the SE RIRs, the polarity-augmented beyerdynamic RIRs $D_1$, the raw Zylia capsule signals $D_2$, and the virtual beamformer outputs $D_3$, totaling approximately 2200 stereo RIRs.

During training, we observed that convolving every source with a full-length RIR produced overly reverberant mixtures. This is expected: PaRIRset RIRs were captured during soundchecks in empty venues, whose reverberation times are substantially longer than those of the same rooms with a full audience~\cite{beranek1960audience}. To compensate, we have further attenuated the tail of each RIR by $\sim \mathcal{U}(0, 1)$.

\section{Experimental Setup}\label{sec:experimental_setup}

With the clean source datasets MUSDB18HQ and MOISESDB, CrowdioSet for audience noise, and PaRIRset for reverberation established, we have proceeded to train a baseline model for Live MSS. We have selected SCNet~\cite{tong2024scnet} as our backbone because it is open-source MSS and shows very competitive performance on MUSDB18HQ.

\subsection{Training Details}\label{subsec:training_details}

We have trained all models from scratch using the default SCNet hyperparameters except for some minor modifications. First, we have reduced the initial learning rate from $5 \times 10^{-4}$ to $4 \times 10^{-4}$ when co-training on MUSDB18HQ and MOISESDB as prescribed \cite{tong2024scnet}. Second, we have removed Exponential Moving Average (EMA), which has simplified our pipeline with negligible impact on performance: our co-trained baseline achieves 9.0\,dB SDR on the MUSDB18HQ test set, same as the 9.0\,dB reported in the SCNet paper~\cite{tong2024scnet} when training on MUSDB18HQ alone and close to the 9.2\,dB reported in their co-training scenario. All models use SCNet's default random time-shift augmentation (up to 2\,s) and have been trained for 200 epochs on a single NVIDIA L40S GPU (approximately 7 days per model).

We use batch sizes of 10 or 8 (with/without audience), split into two groups per batch to enable inter-batch cross-song source remixing~\cite{jeon2024does}. We have verified that remixing does not hurt in our setting: comparing remixing against no remixing yielded a non-significant difference of $+0.07$\,dB SDR on separating noisy mixtures (paired $t$-test, $p=0.53$).

We have trained four models, each corresponding to a different training dataset configuration, which we refer to as \textit{clean}, \textit{rev}, \textit{noisy}, and \textit{noisyrev} respectively. The training datasets for each model are as follows:
\begin{itemize}
  \setlength{\itemsep}{0pt}
  \setlength{\parskip}{0pt} 
    \item \textbf{clean}: MUSDB18HQ + MOISESDB;
    \item \textbf{rev}: (MUSDB18HQ + MOISESDB) $\circledast$ PaRIRset;
    \item \textbf{noisy}: MUSDB18HQ + MOISESDB + CrowdioSet;
    \item \textbf{noisyrev}: (MUSDB18HQ + MOISESDB + CrowdioSet) $\circledast$ PaRIRset.
\end{itemize}

As training difficulty increases with the added degradations, we have found it necessary to adjust the initial learning rate for each condition: $\{4, 3.5, 2, 2\} \times 10^{-4}$ for \textit{clean}, \textit{rev}, \textit{noisy}, and \textit{noisyrev} respectively. The best validation checkpoints have been reached at epochs 198, 199, 197, and 144. Both audience mixing and the RIR convolution ($\circledast$) have been applied within the training loop as described in Sections~\ref{sec:crowdioset} and~\ref{sec:parirset}, so that each epoch presented the model with novel combinations of sources, audiences, and room impulse responses. In reverberant conditions, our objective has been to strictly separate the sources, keeping reverberation in the model's outputs and targets.

\subsection{Evaluation}\label{subsec:evaluation}

For evaluation we have used the MUSDB18HQ test set to keep the evaluation compatible with previous work, using our manually mixed \textit{audience} stems test set and the PaRIRset test set, reducing reverberation ---this time in a deterministic manner: 25\% of samples without reverberation, 25\% with the RIR as it is, and the remaining 50\% split equally between tails attenuated by $-6$\,dB and $-12$\,dB. The same degradations that we have proposed for the training data can be applied to the test data. We have evaluated each model under the four evaluation conditions (\textit{clean}, \textit{rev}, \textit{noisy}, \textit{noisyrev}).

In addition, we have included SAM Audio~\cite{shi2025samaudio} to provide a baseline capable of audience isolation. We have included it in its textual prompting mode, taking the SAM Audio Large pre-trained variant, keeping one re-ranking candidate and splitting the test mixtures with rectangular windows of 20s and no overlap to fit in memory. We have used the text queries \textit{`vocals'}, \textit{`drums'}, \textit{`bass'}, \textit{`other'}, and \textit{`crowd'} (the latter yielding better results than \textit{`audience'}). 

As in SCNet, we have used SDR~\cite{le2019sdr} as the primary objective metric taking non-reverberant or reverberant references depending on the evaluation condition. Since SDR does not always correlate with perceived quality~\cite{guso2022loss, jaffe2025musical}---particularly for diffusion-based models such as SAM Audio, whose non sample-aligned outputs can be penalized by SDR independently of perceptual quality--- we have complemented the objective evaluation with a subjective AB listening test. Given the limited statistical power of small-sample listening tests, we have narrowed the evaluation to two concrete tasks: \textit{vocals isolation} and \textit{audience isolation}. We have selected 10 real concert recordings from the audience, sourced from social media platforms, split into two groups of five. For the vocals isolation task, participants have been presented with a mixture and two separated vocals stems---one from our \textit{noisy} model and one from the \textit{clean} baseline---and asked which separation of the \textit{singer's voice} (as opposed to the crowd) they preferred, if any. For the audience isolation task, participants have compared the audience stem from our \textit{noisy} model against SAM Audio. A total of 26 participants have taken part in the study, comprising media researchers and professional musicians or sound engineers. We have used \cite{zhang_barry_sun_hines_2021} for the online web-based AB listening test interface, randomizing the order of presentation of each model's output.

\section{Results and Discussion}\label{sec:results}
 
Table~\ref{tab:every_source} reports per-source SDR for the most challenging evaluation condition (\textit{noisyrev}) alongside the \textit{clean} results in parentheses. SAM Audio performs substantially worse than all our models in both conditions, achieving only 3.72\,dB on clean \textit{vocals}---well below the 10.05\,dB of our \textit{clean} baseline. We hypothesize that the gap is caused by
the mismatch between the open-vocabulary
nature of SAM Audio and the fixed-taxonomy setting of MUSDB18HQ. We
observe that SAM Audio tends to either extract the source or fail entirely, and successful extractions still yield positive SDR, suggesting the gap is not purely a sample-alignment artifact; nonetheless, we rely primarily on the subjective evaluation.

\begin{table}[t]
  \centering
  \resizebox{\columnwidth}{!}{%
  \renewcommand{\arraystretch}{1.4}
  \begin{tabular}{c|c|c|c|c}
    & vocals & drums & bass & other \\[2pt]
    \hline
    \hline
    clean & $1.46_{\pm 3.4}$ ($\bm{10.05}$) & $7.04_{\pm 3.6}$ ($10.62$) & $4.15_{\pm 3.5}$ ($\bm{8.97}$) & $1.43_{\pm 3.5}$ ($\bm{6.69}$) \\[2pt]

    rev & $1.10_{\pm 3.6}$ ($9.01$) & $\bm{8.23}_{\pm 3.1}$ ($9.86$) & $5.36_{\pm 3.7}$ ($7.77$) & $\bm{3.63}_{\pm 2.4}$ ($5.98$)\\[2pt]

    noisy & $\bm{3.26}_{\pm 2.9}$ ($9.97$) & $7.00_{\pm 3.3}$ ($\bm{10.68}$) & $3.91_{\pm 3.6}$ ($8.85$) & $1.79_{\pm 3.2}$ ($\bm{6.69}$) \\[2pt]

    noisyrev & $3.15_{\pm 3.1}$ ($9.07$) & $8.19_{\pm 3.3}$ ($9.98$) & $\bm{5.66}_{\pm 3.6}$ ($7.86$) & $3.55_{\pm 2.7}$ ($5.88$) \\[2pt]
    
    SAMAudio$^\dagger$ & $-0.45_{\pm 1.9}$ ($3.72$) & $3.30_{\pm 2.6}$ ($5.63$) & $1.30_{\pm 3.0}$ ($5.14$) & $-1.82_{\pm 2.0}$ ($-1.74$)\\[2pt]
  \end{tabular}%
  }
  \caption{Individual sources objective results. Mean SDR $\pm$ standard deviation (dB) for the \textit{noisyrev} evaluation condition and, in parentheses, for the \textit{clean} evaluation condition. $^\dagger$Take SDR with caution, diffusion-based SAM Audio may have sample misalignment.}
  \label{tab:every_source}
\end{table}

Among our four models, no single variant uniformly dominates the \textit{noisyrev} condition. The \textit{noisy} model achieves the best \textit{vocals} SDR (3.26\,dB), while \textit{noisyrev} leads on \textit{bass} (5.66\,dB), and \textit{rev} on \textit{drums} (8.23\,dB) and \textit{other} (3.63\,dB). Interestingly, when evaluating on clean data, the \textit{noisy} model retains near-original performance (9.97\,dB vs.\ 10.05\,dB for vocals), suggesting that audience-noise augmentation alone does not substantially degrade the model's ability to handle studio-quality inputs. We therefore recommend the \textit{noisy} model for vocal extraction when the recording conditions are unknown. Reverberation (\textit{rev} and \textit{noisyrev}) appears especially important for the instrument sources.

Table~\ref{tab:audiences} reports audience isolation SDR. In the reverberant evaluation, \textit{noisy} and \textit{noisyrev} perform nearly identically (1.25 vs.\ 1.56\,dB), but in the non-reverberant case the \textit{noisy} model is substantially superior (2.85 vs.\ 1.55\,dB). Both models outperform SAM Audio by a wide margin ($>$3\,dB). \textit{noisy}'s strong vocal isolation and competitive audience separation supports selecting it for the subjective evaluation.

\begin{table}[b]
  \centering
    \begin{tabular}{c|cc}
    & \multicolumn{2}{c}{audience isolation} \\
    & non-reverberant & reverberant \\
    \hline
    noisy  & $\bm{2.85}_{\pm 1.3}$ & $1.25_{\pm 1.4}$ \\
    noisyrev & $1.55_{\pm 1.0}$ & $\bm{1.56}_{\pm 0.9}$ \\
    SAMAudio$^\dagger$ & $-1.94_{\pm 3.0}$ & $-1.60_{\pm 2.7}$ \\
  \end{tabular}
  \caption{Audience isolation results for the three models capable of it. Mean SDR $\pm$ standard deviation (dB) on the reverberant and/or noisy evaluation conditions. $^\dagger$See Table~\ref{tab:every_source}. $^\dagger$See Table~\ref{tab:every_source}.}
  \label{tab:audiences}
\end{table}

\begin{figure}[t]
  \centering
  \includegraphics[trim={0.1cm 1.0cm 0.1cm 0.1cm}, clip,width=0.99\linewidth]{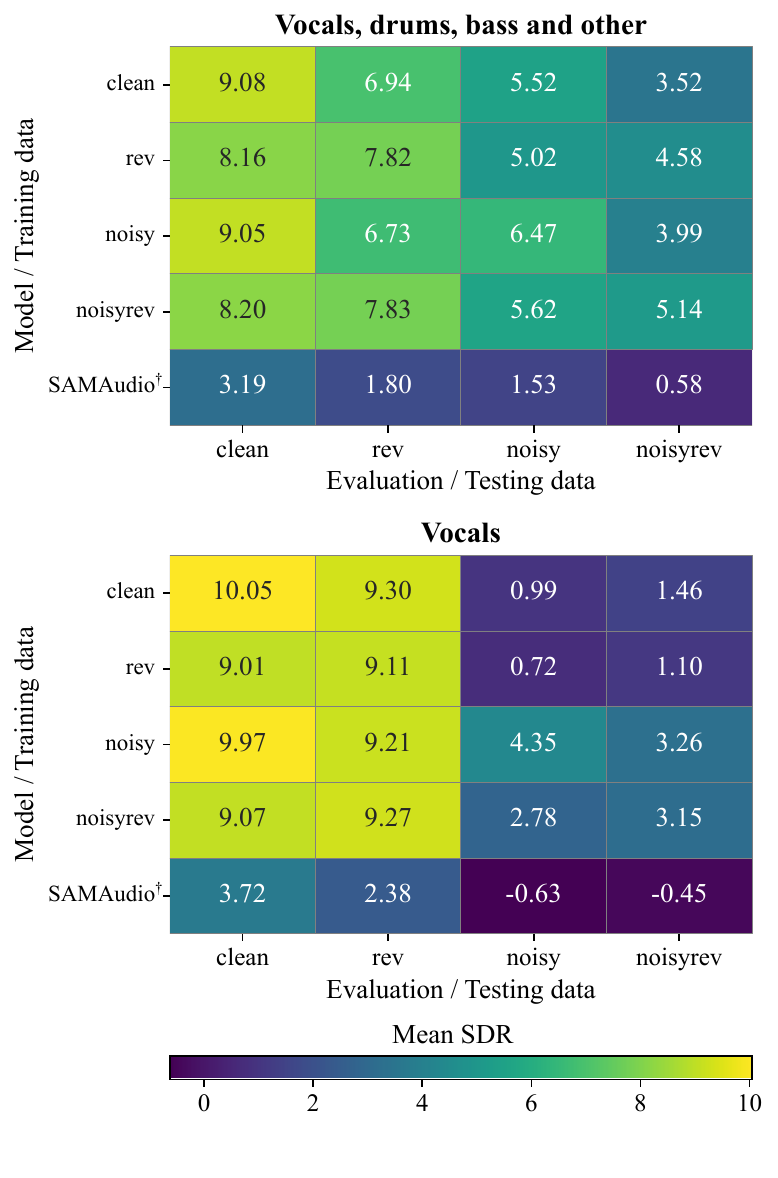}
  \caption{All sources combined (above) and \textit{vocals} isolation (below) mean SDR (dB) results. Each row corresponds to the MSS model while each column corresponds to an evaluation condition. $^\dagger$See Table~\ref{tab:every_source}.}
  \label{fig:heatmap}
\end{figure}

Figure~\ref{fig:heatmap} extends the analysis across all evaluation conditions. The upper heatmap (all sources combined) confirms the expected pattern: in each column (evaluation condition), the best-performing model is the one whose training data matches that condition (except for \textit{rev}, with \textit{noisyrev} marginally ahead). Focusing on the \textit{clean} evaluation condition, the \textit{noisy} model closely follows the \textit{clean} model (9.05 vs.\ 9.08\,dB), suggesting that adding the CrowdioSet \textit{audience} source to training incurs a negligible cost on clean performance. The same happens in the reverberant evaluation condition, where noise in the \textit{noisyrev} model has no cost on performance compared to \textit{rev} (7.83 vs.\ 7.82\,dB). The lower heatmap focuses on \textit{vocals} isolation alone, and shows that for that task the \textit{noisy} model is the most robust across all conditions. Interestingly, when looking at the \textit{rev} evaluation condition, all of our models perform on par. In the \textit{noisyrev} condition, the \textit{noisyrev} model outperforms \textit{noisy} by only 0.39\,dB, while \textit{noisy} gains 1.57\,dB over \textit{noisyrev} in the \textit{noisy} condition ---reinforcing our choice of \textit{noisy} for the subjective evaluation.

\begin{figure}[h]
  \centering
  \includegraphics[trim={1.1cm 0.1cm .0cm .1cm}, clip,width=0.99\linewidth]{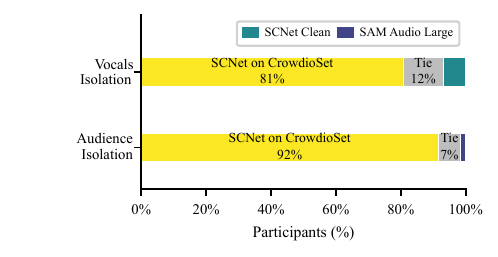}
  \caption{Subjective AB evaluation results comparing our \textit{noisy} model to the baselines: \textit{clean} for \textit{vocals} separation and SAM Audio for \textit{audience} separation.}
  \label{fig:subjective}
\end{figure}

Figure~\ref{fig:subjective} summarizes the AB listening test results. Counting the ties as baseline picks and applying a one-sided binomial test ($H:P>0.5$), listeners significantly preferred our \textit{noisy} model in both tasks: for \textit{vocals} isolation, 105 out of 130 votes favored our model ($p < 10^{-12}$); for \textit{audience} isolation, 119 out of 130 votes favored our model ($p < 10^{-23}$). 
 
\section{Limitations and Future Work}\label{sec:limitations_future}
 
We have not been able to perform a systematic search over the online mixing parameters (probabilities $p_v$, $p_a$, $p_e$ and $g$). The values used in our experiments have been selected through informal listening and may not be optimal. Listening to the model's outputs on real concert recordings, we have observed that while the model separates \textit{ambiences} and \textit{events} effectively---functioning as a denoiser---the \textit{sing-along} component is rarely isolated reliably. We attribute this to two factors. First, the voice conversion approach used to generate sing-alongs produces signals that remain too similar to the original \textit{vocals} stem: some converted voices sing in exact unison with identical phrasing, differing only in timbre, which makes them overly correlated with the lead vocal. Second, the presence of audio effects (reverb and delay) already present on the original \textit{vocals} stems further blurs the boundary between the lead vocal and its synthetic audience counterpart. A future version of CrowdioSet could benefit from using \emph{dry} vocals from \cite{zang2025music} and more expressive synthesis techniques, as well as modeling the artists' reactions to the audience. Finally, the \textit{noisyrev} model converging early at epoch 144 instead of near epoch 200 like the other three models suggests that the combined noise-plus-reverberation augmentation may benefit from architectural modifications or a more thorough hyperparameter search to fully exploit the added training complexity.
 
\section{Conclusions}\label{sec:conclusions}
 
We have presented CrowdioSet, an audience noise dataset, and PaRIRset, a set of impulse responses from 40 concert venues. Together, these extend standard MSS training to the live setting. Experiments with SCNet show that training with CrowdioSet  improves vocal and audience separation on live recordings with minimal degradation on clean studio material, while PaRIRset RIRs yield statistically significant gains over general-purpose RIRs. Objective metrics and a listening test confirm the effectiveness of the approach. Data, samples, code, and weights at: \url{https://enricguso.github.io/crowdioset_parirset}

\section{Acknowledgments}

This work was financially supported by the Catalan Government through the funding grant ACCIO-Eurecat (Project TRAÇA: “IAGen” 2023-2026). We thank the band Cala Vento for allowing us to measure PaRIRset during their \emph{Brindis} tour. We thank the Eurecat team for taking part in the listening test. In particular, we thank Umut Sayın for suggesting the use of the Zylia microphone array, and both him and Joanna Luberadzka for reviewing the manuscript. We also thank Pepe Ferrer from Global Audio Solutions for contributing to PaRIRset with an RIR.

\section{AI Usage Statement}

LLM-based tools (Claude) were used solely to assist with text editing and writing (e.g. wording, grammar, and structure) during the preparation of this manuscript. No AI tools were used to generate, assist with, or review code, experiments, or results.

\bibliography{ISMIRtemplate}

@inproceedings{postolache2023adversarial,
  title={Adversarial permutation invariant training for universal sound separation},
  author={Postolache, Emilian and Pons, Jordi and Pascual, Santiago and Serr{\`a}, Joan},
  booktitle={Proc. IEEE International Conf. on Acoustics, Speech and Signal Processing (ICASSP)},
  year={2023},
}

@inproceedings{eaton2015ace,
  title={The {ACE} {Challenge}—{Corpus} description and performance evaluation},
  author={Eaton, James and Gaubitch, Nikolay D and Moore, Alastair H and Naylor, Patrick A},
  booktitle={Proc. IEEE Workshop on Applications of Signal Processing to Audio and Acoustics (WASPAA)},
  year={2015},
}

@inproceedings{ko2017study,
  title={A study on data augmentation of reverberant speech for robust speech recognition},
  author={Ko, Tom and Peddinti, Vijayaditya and Povey, Daniel and Seltzer, Michael L and Khudanpur, Sanjeev},
  booktitle={Proc. IEEE International Conf. on Acoustics, Speech and Signal Processing (ICASSP)},
  year={2017},
}

@article{traer2016statistics,
  title={Statistics of natural reverberation enable perceptual separation of sound and space},
  author={Traer, James and McDermott, Josh H},
  journal={Proceedings of the National Academy of Sciences},
  volume={113},
  number={48},
  pages={E7856--E7865},
  year={2016},
  publisher={National Academy of Sciences}
}

@inproceedings{anastasopoulou2025freesound,
  title={The {Freesound} {API}: Advances in Audio Search and Retrieval},
  author={Anastasopoulou, Panagiota and Porter, Alastair and Font Corbera, Frederic},
  booktitle={The 9th Web Audio Conf. (WAC)},
  year={2025},
}

@article{mcfee2025librosa,
  title={librosa/librosa: 0.11. 0},
  author={McFee, Brian and McVicar, Matt and Faronbi, Daniel and Roman, Iran and Gover, Matan and Balke, Stefan and Seyfarth, Scott and Malek, Ayoub and Raffel, Colin and Lostanlen, Vincent and others},
  journal={Zenodo},
  year={2025}
}

@inproceedings{farina2000simultaneous,
  title={Simultaneous measurement of impulse response and distortion with a swept-sine technique},
  author={Farina, Angelo},
  booktitle={108th Audio Engineering Society Convention},
  year={2000},
}

@inproceedings{torres2023singer,
  title={Singer Identity Representation Learning using Self-Supervised Techniques},
  author={Torres, Bernardo and Lattner, Stefan and Richard, Gael},
  booktitle={Proc. of the 24th International Society for Music Information Retrieval Conf.},
  year={2023}
}

@misc{rafii2017musdb18,
  author    = {Rafii, Zafar and Liutkus, Antoine and St{\"o}ter, Fabian-Robert and Mimilakis, Stylianos Ioannis and Bittner, Rachel},
  title     = {The {MUSDB18} Corpus for Music Separation},
  year      = {2017},
  url       = {https://doi.org/10.5281/zenodo.1117372}
}

@inproceedings{pereira2023moisesdb,
  title     = {{MoisesDB}: A Dataset for Source Separation Beyond 4-Stems},
  author    = {Pereira, Igor and Ara{\'u}jo, Felipe and Korzeniowski, Filip and Vogl, Richard},
  booktitle = {Proc. of the 24th International Society for Music Information Retrieval Conf.},
  year      = {2023}
}

@inproceedings{rouard2023htdemucs,
  title     = {Hybrid Transformers for Music Source Separation},
  author    = {Rouard, Simon and Massa, Francisco and D{\'e}fossez, Alexandre},
  booktitle = {Proc. IEEE International Conf. on Acoustics, Speech and Signal Processing (ICASSP)},
  year      = {2023},
}

@inproceedings{luo2023bsrnn,
  title     = {Music Source Separation with Band-Split {RNN}},
  author    = {Luo, Yi and Yu, Jianwei},
  booktitle = {Proc. IEEE International Conf. on Acoustics, Speech and Signal Processing (ICASSP)},
  year      = {2023},
}

@inproceedings{tong2024scnet,
  title     = {{SCNet}: Sparse Compression Network for Music Source Separation},
  author    = {Tong, Weinan and Zhu, Jiaxu and Chen, Jun and Kang, Shiyin and Jiang, Tao and Li, Yang and Wu, Zhiyong and Meng, Helen},
  booktitle = {Proc. IEEE International Conf. on Acoustics, Speech and Signal Processing (ICASSP)},
  year      = {2024},
}

@inproceedings{le2019sdr,
  title={{SDR}--half-baked or well done?},
  author={Le Roux, Jonathan and Wisdom, Scott and Erdogan, Hakan and Hershey, John R},
  booktitle={Proc. IEEE International Conf. on Acoustics, Speech and Signal Processing (ICASSP)},
  year={2019},
}

@article{fabbro2024sdx23,
  title   = {The Sound Demixing Challenge 2023 -- Music Demixing Track},
  author  = {Fabbro, Giorgio and Uhlich, Stefan and Lai, Chieh-Hsin and Choi, Woosung and Mart{\'i}nez-Ram{\'i}rez, Marco and Liao, Weihsiang and others},
  journal = {Transactions of the International Society for Music Information Retrieval},
  volume  = {7},
  number  = {1},
  year    = {2024},
}

@article{zang2026msr,
  title   = {Summary of The Inaugural Music Source Restoration Challenge},
  author  = {Zang, Yongyi and Hai, Jiarui and Ge, Wanying and Kong, Qiuqiang and Dai, Zheqi and Wang, Helin and Mitsufuji, Yuki and Plumbley, Mark D.},
  journal = {arXiv preprint arXiv:2601.04343},
  year    = {2026},
}

@inproceedings{plaja2025generating,
  title={Generating Separated Singing Vocals Using a Diffusion Model Conditioned on Music Mixtures},
  author={Plaja-Roglans, Gen{\'\i}s and Hung, Yun-Ning and Serra, Xavier and Pereira, Igor},
  booktitle={Proc. IEEE Workshop on Applications of Signal Processing to Audio and Acoustics (WASPAA)},
  year={2025},
}

@inproceedings{plaja2023carnatic,
  title={Carnatic Singing Voice Separation Using Cold Diffusion on Training Data With Bleeding},
  author={Plaja-Roglans, Gen{\'\i}s and Miron, Marius and Shankar, Adithi and Serra, Xavier},
  booktitle={Proc. of the 24th International Society for Music Information Retrieval Conf.},
  year={2023}
}

@article{mitsufuji2022music,
  title={Music demixing challenge 2021},
  author={Mitsufuji, Yuki and Fabbro, Giorgio and Uhlich, Stefan and St{\"o}ter, Fabian-Robert and D{\'e}fossez, Alexandre and Kim, Minseok and Choi, Woosung and Yu, Chin-Yun and Cheuk, Kin-Wai},
  journal={Frontiers in Signal Processing},
  volume={1},
  year={2022},
  publisher={Frontiers Media SA}
}

@inproceedings{hung2025moises,
  title={Moises-Light: Resource-efficient Band-split U-Net For Music Source Separation},
  author={Hung, Yun-Ning Amy and Pereira, Igor and Korzeniowski, Filip},
  booktitle={Proc. IEEE Workshop on Applications of Signal Processing to Audio and Acoustics (WASPAA)},
  year={2025},
}

@inproceedings{venkatesh2024real,
  title={Real-time low-latency music source separation using hybrid spectrogram-tasnet},
  author={Venkatesh, Satvik and Benilov, Arthur and Coleman, Philip and Roskam, Frederic},
  booktitle={Proc. IEEE International Conf. on Acoustics, Speech and Signal Processing (ICASSP)},
  year={2024},
}

@article{roa2025first,
  title={The first Cadenza challenges: using machine learning competitions to improve music for listeners with a hearing loss},
  author={Roa-Dabike, Gerardo and Akeroyd, Michael A and Bannister, Scott and Barker, Jon P and Cox, Trevor J and Fazenda, Bruno and Firth, Jennifer and Graetzer, Simone and Greasley, Alinka and Vos, Rebecca R and others},
  journal={IEEE Open Journal of Signal Processing},
  year={2025},
  publisher={IEEE}
}

@article{wichern2019wham,
  title={{WHAM!}: Extending Speech Separation to Noisy Environments},
  author={Wichern, Gordon and Antognini, Joe and Flynn, Michael and Zhu, Licheng Richard and McQuinn, Emmett and Crow, Dwight and Manilow, Ethan and Le Roux, Jonathan},
  journal={Proc. Interspeech},
  year={2019},
  publisher={ISCA}
}

@inproceedings{maciejewski2020whamr,
  title={{WHAMR!}: Noisy and reverberant single-channel speech separation},
  author={Maciejewski, Matthew and Wichern, Gordon and McQuinn, Emmett and Le Roux, Jonathan},
  booktitle={Proc. IEEE International Conf. on Acoustics, Speech and Signal Processing (ICASSP)},
  year={2020},
}

@inproceedings{bai2026hq,
  title={{HQ}-{SVC}: Towards high-quality zero-shot singing voice conversion in low-resource scenarios},
  author={Bai, Bingsong and Geng, Yizhong and Wang, Fengping and Wang, Cong and Guo, Puyuan and Gao, Yingming and Li, Ya},
  booktitle={Proc. of the AAAI Conf. on Artificial Intelligence},
  year={2026}
}

@article{beranek1960audience,
  title={Audience and seat absorption in large halls},
  author={Beranek, Leo L},
  journal={The journal of the Acoustical Society of America},
  volume={32},
  number={6},
  year={1960},
  publisher={Acoustical Society of America}
}

@article{zhang_barry_sun_hines_2021, title={{Go Listen}: An End-to-End Online Listening Test Platform}, journal={Journal of Open Research Software}, author={Dan Barry and Qijian Zhang and Pheobe Wenyi Sun and Andrew Hines}, year={2021}}

@inproceedings{jeon2024does,
  title={Why does music source separation benefit from cacophony?},
  author={Jeon, Chang-Bin and Wichern, Gordon and Germain, Fran{\c{c}}ois G and Le Roux, Jonathan},
  booktitle={Proc. IEEE International Conf. on Acoustics, Speech and Signal Processing Workshops (ICASSPW)},
  year={2024},
}

@inproceedings{sheridan2025perceptual,
  title     = {Perceptual Errors in Music Source Separation: Looking Beyond {SDR} Averages},
  author    = {Sheridan, Sean and Benetos, Emmanouil},
  booktitle = {Proc. of the 26th International Society for Music Information Retrieval Conf.},
  year      = {2025},
}

@inproceedings{jaffe2025musical,
  title={Musical source separation bake-off: Comparing objective metrics with human perception},
  author={Jaffe, Noah and Burgoyne, John Ashley},
  booktitle={Proc. IEEE Workshop on Applications of Signal Processing to Audio and Acoustics (WASPAA)},
  year={2025},
}

@article{shi2025samaudio,
  title   = {{SAM} Audio: Segment Anything in Audio},
  author  = {Shi, Bowen and Tjandra, Andros and Hoffman, John and Wang, Helin and Wu, Yi-Chiao and Gao, Luya and Richter, Julius and Le, Matt and Vyas, Apoorv and Chen, Sanyuan and Feichtenhofer, Christoph and Doll{\'a}r, Piotr and Hsu, Wei-Ning and Lee, Ann},
  journal = {arXiv preprint arXiv:2512.18099},
  year    = {2025},
}

@inproceedings{kandpal2022musicenhancement,
  title     = {Music Enhancement via Image Translation and Vocoding},
  author    = {Kandpal, Nikhil and Nieto, Oriol and Jin, Zeyu},
  booktitle = {Proc. IEEE International Conf. on Acoustics, Speech and Signal Processing (ICASSP)},
  year      = {2022},
}

@inproceedings{guso2022loss,
  title={On loss functions and evaluation metrics for music source separation},
  author={Gus{\'o}, Enric and Pons, Jordi and Pascual, Santiago and Serr{\`a}, Joan},
  booktitle={Proc. IEEE International Conf. on Acoustics, Speech and Signal Processing (ICASSP)},
  year={2022},
  organization={IEEE}
}

@article{stoter2019open,
  title={Open-unmix - A reference implementation for music source separation},
  author={St{\"o}ter, Fabian-Robert and Uhlich, Stefan and Liutkus, Antoine and Mitsufuji, Yuki},
  journal={Journal of Open Source Software},
  volume={4},
  number={41},
  pages={1667},
  year={2019}
}

@inproceedings{zang2025music,
  title={Music source restoration},
  author={Zang, Yongyi and Dai, Zheqi and Plumbley, Mark D and Kong, Qiuqiang},
  booktitle={2025 IEEE International Workshop on Multimedia Signal Processing (MMSP)},
  pages={138--143},
  year={2025},
  organization={IEEE}
}

\end{document}